# Learning Recommendations While Influencing Interests

Rahul Meshram, D. Manjunath and Nikhil Karamchandani

*Abstract*— **Personalized recommendation systems (RS) are extensively used in many services. Many of these are based on learning algorithms where the RS uses the recommendation history and the user response to learn an optimal strategy. Further, these algorithms are based on the assumption that the user interests are rigid. Specifically, they do not account for the effect of learning strategy on the evolution of the user interests. In this paper we develop influence models for a learning algorithm that is used to optimally recommend websites to web users.**

**We adapt the model of [1] to include an item-dependent reward to the RS from the suggestions that are accepted by the user. For this we first develop a static optimization scheme when all the parameters are known. Next we develop a stochastic approximation based learning scheme for the RS to learn the optimal strategy when the user profiles are not known. Finally, we describe several user-influence models for the learning algorithm and analyze their effect on the steady user interests and on the steady state optimal strategy as compared to that when the users are not influenced.**

## I. Introduction

Personalised recommendation systems (RS) are being used in a large variety of situations, e.g., suggesting movies and music on streaming services, YouTube videos, items to shop, and news articles to read. Many of these are learning systems that use the history of recommendations and user reactions to determine the user interest profile and hence the sequence of recommendations. An implicit assumption in the design of these systems is that the user interests are 'rigid,' i.e., they are not influenced by the recommendation history. In this paper, we analyze a recommendation system in which while learning to recommend optimally, the learning process, i.e., the sequence of recommendations, influences the user interests which now evolve with time. We believe that this is an important aspect of recommendation systems that needs to be better understood.

Our starting point in this paper is the learning algorithm described for a personalised RS in [1]. This RS considered a system in which the objects are complex in that the liking of the item depends on the random set of 'sub-items' that are present when the item is shown. This in turn was motivated by recommendations from the `Stumbleupon` website[1]. We introduce some modifications to their system to model the influence of the learning process on the user interests. This model is also convenient to develop a tractable influence model, albeit a stylized one.

The system has a number of topics that may interest a user at any time and there are a number of websites, each of which cover only a subset of these topics at any time. A user has an interest profile, unknown to the RS, that describes the level of interest in each topic. A website is characterized by the frequency with which each of topic is covered. The RS makes a sequence of suggestions of websites till the user chooses ('likes') one, i.e., the recommended site has the topic of interest. The RS experiences a cost that is increasing in the number of trials before a suggestion is accepted. Our first key departure from the model of [1] is in our assumption that the RS receives a website-dependent reward from the site that the user likes. Thus the RS has to use a recommendation strategy that maximizes its payoff, i.e., reward minus cost. The reward skews the recommendation sequence towards profitable websites and motivates the modeling of the learning process influencing the user interest towards topics that are covered by profitable sites.

The rest of the paper is organized as follows. We first consider the system in which the characteristics of both the user and the websites are known and fixed, and the RS needs to determine the optimal randomized recommendation strategy. This is the subject of Section II. In Section III we develop the learning algorithm and its analysis when the user characterization is not available to the RS. The optimum strategy is learnt through a binary feedback from the user and is along the lines of [1] but modified for the new payoff structure. In Section IV we describe models for the influence of the recommendation sequence on the user's interest and then study the effect of the interaction between the learning algorithm of the RS and the changing interest profile of the user.

### A. Literature Survey

The literature on recommendation systems in varied and extensive. Personalised RS are typically based on a form of collaborative filtering (suggest items that are preferred by other 'similar' users), or are content-based (suggestions are derived from history of user behavior), or a combination of both. See [2], [3] for excellent surveys. Detailed descriptions of personalized news recommendations are available in [1], [4]–[8]. Collaborative filtering and the probabilistic clustering method is used in [4]. Online learning algorithms are described in [1]. A hybrid filtering technique is considered in [5] using a Bayesian learning model. Context-based filtering techniques are used in [5], [7]. As we mentioned earlier, all

Rahul Meshram is with the Electrical and Computer Engineering Department at Waterloo University, Canada. D. Manjunath and Nikhil Karamchandani are with the Electrical Engineering Department of IIT Bombay.

[1]`https://www.stumbleupon.com/`

of these assume that the user interest is rigid and does not evolve. In fact, the stability of user interest has been studied in [5] via a large scale log analysis of Google News users. User interest prediction are of interest in [5], [9]. But these do not study, or model, the cause of the change of user interests. In this work, we take a first step towards filling this gap.

## II. Model, Notation, and the Static Problem

Consider a RS that has a set of $M$ websites. There are a total of $N$ topics that these websites can cover. Each website will cover one or more of these topics. We will assume that when website $m$ is recommended, it will have covered topic $n$ with probability $p_{m,n}$ independent of everything else. The row vector $p_m = [p_{m,1}, \ldots, p_{m,N}]$ will be called the *publishing strategy* of website $m$. The matrix $P = [[p_{m,n}]]$ will be called the *publishing matrix*. $\sum_{n=1}^{N} p_{m,n}$ is the expected number of topics that are covered by website $m$ on any showing. Further, $\sum_m p_{m,n}$ is the expected number of websites that cover topic $n$ and we assume this to be strictly positive, i.e., we assume that there is at least one website that covers each topic $n$.

When a user begins its search for a website, we say that a user session has begun. At the start of the session the user is interested in topic $n$ with probability $\theta_n$ and we assume that this remains unchanged during the session. $\theta := [\theta_n]$, a row vector, will be called the *interest profile* of the user. Clearly, $\sum_{n=1}^{N} \theta_n = 1$. We assume that the RS uses a randomized strategy and recommends website $m$ independently of everything else, with probability $x_m$ in each trial; $x := [x_m]$, a row vector, is called the strategy of the RS; also $\sum_{m=1}^{M} x_m = 1$. The user can either accept the recommendation and click through to the website. Or it can reject the recommendation and ask for another recommendation and the RS obliges. This continues till the recommendation is accepted by the user at which point the session ends. Each recommendation in a session will be called a trial. Clearly, the number of trials in a session is random and depends on $x$, $\theta$, and $P$. We will assume that the recommendations in each trial of a session are independent and drawn from the same distribution $x$.

For each click-through, website $m$ offers a fixed revenue $r_m$ to the RS. Let $r := [r_m]$, a row vector, be the revenue vector. The user experience in a session depends on the number of trials required to obtain a topic of interest; the smaller the number of trials the better the user experience. To capture this, we define a cost function for each session, $c_l$ where $l$ is the number of trials in the session. We will assume that $c_l$ is convex and non decreasing in $l$. Associating such a cost for each session ensures that RS does not recommend too many low interest sites in a session even though they may fetch a higher revenue. In the rest of this paper, we assume that the cost function is polynomial in $l$, specifically, we let $c_l = l^{\kappa}$, for a finite $\kappa$. Thus the objective of the RS is to devise a strategy $x$ that maximises the net expected reward, revenue minus the cost, in each session.

For a fixed, $\theta$, $x$, $r$, and $P$, we now calculate the expected revenue per session. Given that the user interest is topic $n$ in a session, let $\rho_n$ denote the conditional probability of accepting a recommendation in a trial. Clearly, this is given by

$$\rho_n := \sum_{m=1}^{M} x_m p_{m,n}. \tag{1}$$

Given that the interest is in topic $n$, the conditional probability that the session ends with a click-through to site $m$ is given by

$$\frac{x_m p_{m,n}}{\sum_{m=1}^{M} x_m p_{m,n}}.$$

The expected reward from a session under a recommendation strategy $x$, say $R(x)$, is thus

$$R(x) = \sum_{n=1}^{N} \theta_n \sum_{m=1}^{M} r_m \left[ \frac{x_m p_{m,n}}{\sum_{m_1=1}^{M} x_{m_1} p_{m_1,n}} \right].$$

Rearranging the above, we obtain

$$R(x) = \sum_{n=1}^{N} \theta_n \left[ \frac{\sum_{m=1}^{M} r_m x_m p_{m,n}}{\sum_{m_1=1}^{M} x_{m_1} p_{m_1,n}} \right]. \tag{2}$$

Let $T$ be the random variable denoting the number of trials in a session. Given $n$, $T$ has a geometric distribution with parameter $\rho_n$. Thus

$$\Pr(T = l) = \sum_{n=1}^{N} \theta_n \rho_n (1 - \rho_n)^{(l-1)}$$

for $l \geq 1$. The expected cost under a recommendation strategy $x$, say $C(x)$, in a session is

$$C(x) = \sum_{n=1}^{N} \theta_n \left( \sum_{l=1}^{\infty} c_l \ \rho_n (1 - \rho_n)^{l-1} \right).$$

We assume that $x_m \geq \epsilon > 0$, i.e., all websites are recommended with a non zero probability. $\epsilon$ can be seen to be a guarantee from the RS to the websites. Since $\sum_{m=1}^{M} p_{m,n} > 0$, and $x_m \geq \epsilon$, we see that $\rho_n$ is strictly positive, i.e., $\rho_n \geq \delta > 0$. This ensures that the expected cost $C(x)$ in each session is finite. We can rewrite the expected cost $C(x)$ in the following way.

$$C(x) = \sum_{n=1}^{N} \theta_n \left[ \sum_{l=1}^{\infty} \Delta c_l (1 - \rho_n)^{(l-1)} \right] \tag{3}$$

where,

$$\Delta c_l = \begin{cases} c_1 & \text{if } l = 1, \\ c_l - c_{l-1} & \text{if } l > 1. \end{cases} \tag{4}$$

Derivation of Eqn. (3) is given in Appendix A.

*Lemma 1:* $C(x)$ is convex in $x$.

*Proof:* Proof is given Appendix B. ■

The expected reward per session is thus $(R(x) - C(x))$ and the optimal strategy, $x^*$, is the solution to the following optimization problem.

$$\begin{aligned} \max_x \quad & F(x) = R(x) - C(x) \\ \text{s.t.} \quad & \textstyle\sum_{m=1}^{M} x_m = 1, \\ & x_m \geq \epsilon \ \text{ for } m = 1, 2, \ldots, M. \end{aligned} \tag{P1}$$

Observe that $R(x)$ is a convex combination of ratios of linear functions of $x$. Thus (P1) is a fractional program [10], [11] that is the sum of two terms—a convex function of $x$ and a sum-of-ratios term. In [11] it is shown that (P1) is essentially NP-complete and an interior point method to obtain an $\epsilon$-optimal solution is described. Our interest here is not in the efficient solution of (P1). Instead, it is on the effect of the reward structure on the recommendations. We study this via some numerical examples below.

### A. Numerical Examples

We now present some numerical examples when $P$, $\theta$, and $r$ are known to the RS. Since the objective is to illustrate, we consider a small system with five websites and four topics, i.e., $M = 5$, $N = 4$. We will use the following publishing matrix in all the examples of the paper.

$$P = \begin{bmatrix} 0.25 & 0.25 & 0.25 & 0.25 \\ 0.25 & 0.25 & 0.25 & 0.25 \\ 0.10 & 0.40 & 0.45 & 0.05 \\ 0.10 & 0.05 & 0.15 & 0.70 \\ 0.65 & 0.10 & 0.20 & 0.05 \end{bmatrix}.$$

The publishing matrix $P$ is chosen to have some variations in the coverage of topics. Specifically, websites 1 and 2 have equal coverage, website 3 covers topics 2 and 3 approximately equal and has negligible coverage of other topics, websites 4 and 5 are more specialised in that they cover, respectively, topics 4 and 1 with significantly higher frequency than others.

Different examples will have different reward structures for the RS. In the following, we determine the optimal strategy $x^*$ by solving (P1) as a static optimization problem in MATLAB. We also use $\epsilon = 0.01$, and $\kappa = 2.5$, i.e., $c_l = l^{2.5}$.

In the first example, we consider reward $r = [150, 125, 150, 400, 100]$ and user profile $\theta = [0.03, 0.05, 0.02, 0.9]$. Note that reward from website 4 is the highest compared to other websites and that website covers topic 4 most frequently. Also, from $\theta$, note that the user interest is high for topic 4. The optimal recommendation strategy $x^*$ and net revenue $F(x^*)$ are shown in Table I. As expected, site 4 is recommended most frequently because it covers topic 4 that matches the user interest.

TABLE I

FOR $r = [150, 125, 150, 400, 100]$ AND $\theta = [0.03, 0.05, 0.02, 0.9]$, WE SEE THE EFFECT OF RECOMMENDATION STRATEGY ON NET REVENUE.

| | Recommendation strategy | Net revenue $F(x) = R(x) - C(x)$ |
|---|---|---|
| 1 | $x^* = [0.01, 0.01, 0.01, 0.96, 0.01]$ | 336.20 |
| 2 | $x^* = [0.01, 0.01, 0.20, 0.76, 0.01]$ | 337 |
| 3 | $x^* = [0.01, 0.01, 0.25, 0.72, 0.01]$ | 337 |

In the second example, we use $r = [400, 125, 150, 150, 100]$ and $\theta = [0.03, 0.05, 0.02, 0.9]$. After solving (P1), we obtain the recommendation strategy $x^* = [0.96, 0.01, 0.01, 0.01, 0.01]$ and the net revenue $F(x^*) = 305.33$. We observe that the website 1 is the highly recommended website compared to other websites. We further note that the net revenue obtained in this example is less than that of first example. This is because the highest rewarding website 1 covers all topics with equal probability, while the user interest is highly skewed toward topic 4.

In the next section we consider the case when the user interest $\theta$ is unknown and we describe a learning algorithm that learns the optimum recommendation strategy based on the feedback from each session.

## III. An Online Learning Algorithm

We now consider the case when the publishing matrix $P$ and the reward vector $r$ are known but the user interest $\theta$ is not known. The algorithms and the corresponding theory that we develop here can be extended to the case when either one of $P$ or $r$ or both are also unknown. Assuming this knowledge simplifies the pedagogy.

Recall from the preceding section that in a session each recommendation is a trial and the RS sequentially recommends websites until the user accepts a recommendation; the accepted site was the first one to show the topic that was of interest to the user in that session. We assume that information about the accepted website and the topic of interest is available to the RS. Further, the number of trials in the session is also known to the RS. Let $x(s)$ be the strategy at the beginning of session $s$, $T(s)$ the number of trials required to achieve a click-through, $m(s)$ the accepted website and $n(s)$ the topic that achieved the click through in the session. We reiterate that we have assumed that $\sum_{m=1}^{M} p_{m,n} > 0$ and $x_m(s) \geq \epsilon > 0$ for all $m, s$, and hence $\rho_n \geq \delta > 0$.

The learning algorithm is stochastic gradient based where the actual gradient is not available to update the strategy; instead a noisy version of the gradient is known after each session based on the feedback from the user. The analysis uses the stochastic approximation approach.

We first describe the intuition for adaptation of the recommendation strategy and then describe and analyze the algorithm. For intuition, we look at the gradients of $R(x)$ and $C(x)$ w.r.t. $x$. Taking partial derivative of $R(x)$ and $C(x)$ w.r.t. $x_m$ from (2) and (3), and using the definition of $\Delta c_l$ from (4), we obtain the following.

$$\frac{\partial R(x)}{\partial x_m} = \sum_{n=1}^{N} \frac{\theta_n p_{m,n}}{\rho_n} \left[ r_m - \sum_{k=1}^{M} \frac{x_k p_{k,n}}{\rho_n} r_k \right],$$

$$\frac{\partial C(x)}{\partial x_m} = -\sum_{n=1}^{N} \theta_n p_{m,n} \sum_{l=1}^{\infty} l \Delta c_{l+1} (1-\rho_n)^{l-1}. \quad (5)$$

The details of the derivation of Eqn. (5) are given in Appendix C.

Now consider session $s$ in which the user has accepted website $m(s)$ and clicked on topic $n(s)$. Observe from Eqn. (5) that the term $\sum_{k=1}^{M} \frac{x_k(s) p_{k,n(s)}}{\rho_{n(s)}} r_k$ is the conditional expected reward given that the user has performed a click-through on topic $n(s)$.

The term $\left(r_{m(s)} - \sum_{k=1}^{M} \frac{x_k(s) p_{k,n(s)}}{\rho_{n(s)}} r_k\right)$ is the difference between the reward that is obtained from website $m$ and the conditional expected reward given that the user accepted site $m(s)$ and topic $n(s)$. Since $m(s)$ and $n(s)$ are random, this term also represents random quantity. Further, since the number of trials in a session is random, $l\Delta c_{l+1}$ is also random.

The above discussion motivates us to consider feedback to the RS in the following form. Define

$$g_m(s) \quad := \quad \begin{cases} \frac{1}{x_{m(s)}}\Delta & \text{if } m(s) = m, \text{ and } n(s) = n \\ 0 & \text{otherwise.} \end{cases} \quad (6)$$

$$\Delta = \left[ r_m - \frac{1}{\rho_n(s)} \sum_{k=1}^{M} r_k x_k(s) p_{k,n} + T(s)\Delta c_{T(s)+1} \right]$$

Define $G(s) := [g_1(s), \cdots, g_M(s)]$ and use the following update to the strategy at the beginning of session $s+1$.

$$x(s+1) = \Pi_{\mathcal{D}_\epsilon}[x(s) + a(s)G(s)]. \quad (7)$$

Here, $a(s)$ is the learning rate parameter that satisfies $a(s) \to 0$ as $s \to \infty$, $\sum_{s=1}^{\infty} a(s) = \infty$, and $\sum_{s=1}^{\infty} a^2(s) < \infty$, and $\mathcal{D}_\epsilon = \left\{ x : x = [x_1, \cdots, x_M]^T, \sum_{m=1}^{M} x_m = 1, \ x_m \geq \epsilon \right\}$, $0 \leq \epsilon \leq \frac{1}{M}$. $\Pi_{\mathcal{D}_\epsilon}$ is the Euclidean projection and defined as follows.

$$\Pi_{\mathcal{D}_\epsilon}[y] := \arg\min_{x \in \mathcal{D}_\epsilon} ||y - x||_2 \qquad x \in \mathbb{R}^M.$$

We remark that the learning algorithm defined by Eqns. (6) and (7) can be viewed as a projected stochastic gradient algorithm.

We first want to show that the adaptation term $G(s)$ is a noisy version of the gradient of the objective function $F(x) := R(x) - C(x)$ and that the noise has zero mean and bounded variance. To show this, we first prove that $\mathrm{F}(x)$ is continuously differentiable and Lipschitz in the following lemma.

*Lemma 2:* The objective function is $\mathrm{F}(x) = C(x) - R(x)$ and $\mathrm{F} : \mathcal{D}_\epsilon \to \mathbb{R}$ is continuously differentiable and Lipschitz. i.e.,

$$\nabla \mathrm{F}(x) = \nabla R(x) - \nabla C(x),$$

where $\nabla \mathrm{F}(x) = \left[\frac{\partial \mathrm{F}(x)}{\partial x_1}, \cdots, \frac{\partial \mathrm{F}(x)}{\partial x_M}\right]$, $\nabla R(x) = \left[\frac{\partial \mathrm{R}(x)}{\partial x_1}, \cdots \frac{\partial \mathrm{R}(x)}{\partial x_M}\right]$, $\nabla C(x) = \left[\frac{\partial \mathrm{C}(x)}{\partial x_1}, \cdots \frac{\partial \mathrm{C}(x)}{\partial x_M}\right]$ and $||\mathrm{F}(\tilde{x}) - \mathrm{F}(\hat{x})|| \leq \tilde{L}||\tilde{x} - \hat{x}||$, for all $\tilde{x}, \hat{x} \in \mathcal{D}_\epsilon$, and $\tilde{L}$ is a Lipschitz constant.

*Proof:* See Appendix D ■

We are now ready to show that the update strategy $G(s)$ is a noisy version of the gradient where the noise is zero mean with bounded variance.

*Lemma 3:* $G(s)$ is a noisy version of gradient of $\mathrm{F}(x)$, that is, $G(s) = \nabla \mathrm{F}(x(s) + V(x(s))$, where $V(x(s))$ is random variable such that $\mathbb{E}[V(x(s))] = 0$ and $\mathbb{E}[||V(x(s)||^2] < \infty$.

*Proof:* See Appendix F. ■

In Theorem 1 below, we show that the continuous interpolated process of update algorithm of (7) asymptotically tracks the solution of the limiting o.d.e.

$$\dot{x}(s) = \nabla \mathrm{F}(x(s)) + z. \quad (8)$$

Here, $z$ is an error due to projection.

*Theorem 1:* Let $a(s)$ be non negative sequence such that $a(s) \to 0$ as $s \to \infty$, $\sum_{s=1}^{\infty} a(s) = \infty$ and $\sum_{s=1}^{\infty} (a(s))^2 < \infty$. Then, the iterates $x(s)$ generated by the update equation (7) converges to $x^*$ almost surely. Here, $x^* \in B := \{x : \nabla F(x) + z = 0\}$ is the set of all stationary points. Also, $\lim_{s\to\infty} F(x(s)) = F(x^*)$.

*Proof:* We present a sketch of the proof. From Lemmas 2 and 3, we note that the the Algorithm (7) with decreasing stepsizes $a(s)$ satisfies the assumptions **A**1, **A**2, **A**3 and **A**4 in [12, Chapter 2, Section 2.1]. This allows us to construct a piecewise linear and continuous interpolated process of iterate (7) and show that the interpolated process tracks the solution of o.d.e. (8). Further, the solution of o.d.e. converges to the limit point set of the o.d.e., i.e., to the set of all stationary points, $B = \{x : \nabla F(x) + z = 0\}$. Hence, the Algorithm (7) converges to a stationary point in $B$. The detailed proof is given in [12, Chapter 2, Chapter 5, Section 5.4], [13, Chapter 5, Section 5.2]. ■

*Remark 1:* In general, for the learning algorithm to converge to the solution of the o.d.e., the algorithm has to follow the stability property, i.e., the algorithm should converge to a bounded value. Often, in stochastic approximation problems, the stability of the algorithm is assumed. But in our problem, the stability issue of the iterates (7) is solved by the projection operation onto to the bounded convex set $\mathcal{D}_\epsilon$. Hence we do not need to make any assumption about the stability property of the algorithm.

We next study the convergence of the Algorithm (7) via some numerical examples.

### *A. Numerical examples*

We simulate an RS where websites are recommended in session $s$ using strategy $x(s)$. We adapt $x(s)$ at the beginning of session $s+1$ using Eqn. (7). We use $a(s) = \frac{0.01}{(1+s^{2/3})}$. The system is also randomly initialized, i.e., $x(1)$ is randomly drawn from the uniform distribution on $[0, 1]$ We use publishing matrix $P$ from Section II-A.

In Fig. 1, we illustrate the convergence of Algorithm (7) for the first example of Section II-A. We plot $x_m(s)$ for website $m$ for $m = 4, 5$. We notice in Fig. 1-a that $x_4(s) \to 0.96$ and $x_5(s) \to 0.01$ as $s$ increases. For this example, from Table I, we know that $x^* = [0.01, 0.01, 0.01, 0.96, 0.01]$. Thus $x_4(s) \to x_4^*$, $x_5(s) \to x_5^*$. Similarly other $x_m(s) \to x_m^*$. Note that the convergence rate of $x_m(s)$ in this algorithm is determined by stepsizes $a(s)$. We also plot the net revenue $F(x(s))$ as function of $s$ in Fig 1-b. We notice that $F(x(s))$ also converges to $F(x^*)$. In Fig. 1-c, the error function $||x(s) - x^*||$ verses $s$ is plotted. We observe that it decreases very quickly.

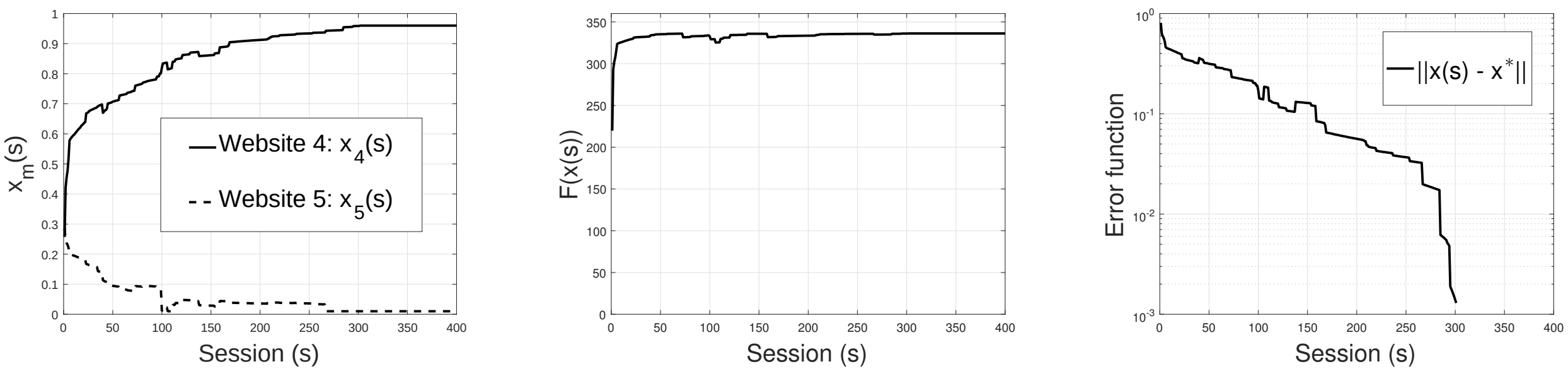

a) b) c)

Fig. 1. Plot (a) illustrates the recommendation strategy $x_m(s)$ for website $m$ verses session $s$. Plot (b) describes the objective function $F$ verses session $s$. In plot (c), we describe the error function $||x(s) - x^*||$ as function of $s$. It is illustrated for $r = [150, 125, 150, 400, 100]$ and $\theta = [0.03, 0.05, 0.02, 0.9]$.

## IV. On Recommendations Influencing User Interests

In the preceding section we studied online learning for users whose interest for topics is unknown but it is static. It may be that a user's intrinsic interest in the various topics is influenced by the recommendations from the RS and hence is dynamic. This changing of the user interest in turn also affects the recommendation strategy of RS, i.e., there is an interaction between $x(s)$ of the RS and the $\theta(s)$ of the user. In this paper we assume that the RS is learning assuming that the user interests are static and it does not know that it is influencing the user's interests. Our interest in this section is to model such an interaction and study its effects. As in the previous section, here we also assume that $P$ and $r$ are known to RS.

Recall that $x(s)$ is updated at the beginning of session $s+1$ according to Eqns. (6) and (7). Since user interest is changing, its value at the beginning of session $s$ is denoted by $\theta(s)$. Let $\chi_n(s)$ denote the random variable corresponding to the number of times topic $n$ is seen during session $s$. To model the influence, we introduce the following two parameters. $\xi_n$, $0 \leq \xi_n \leq 1$, denotes the 'intrinsic' interest in topic $n$, $\beta_n$, $\beta_n \geq 1$, is the influence factor on topic $n$, and $v_n(s)$ is the influence level in session $s$. We consider two kinds of influence models and define $v_n(s)$ differently for these types.

1) *Type-A* influence where the users are negatively influenced by repeated showing of a topic and interest for the topic decreases with increased showings. For such a topic, the influence level from session $s$ is modeled as follows.

$$v_n(s) = \xi_n \beta_n^{-\chi_n(s)}. \tag{9}$$

2) *Type-B* influence where the user interest increases for topics that are shown frequently. For a topic that is influenced in this manner, the level from session $s$ is modeled as follows.

$$v_n(s) = \xi_n (1 - \beta_n^{-\chi_n(s)}). \tag{10}$$

For both types of influence, $\beta_n = 1$, corresponds to the case when there is no influence. For a given user, different topics could be influenced differently and we reiterate that influence type is associated with the topic. The effect of $v_n(s)$ on $\theta(s+1)$ is modeled as below.

$$w_n(s) := \frac{v_n(s)}{\sum_{n=1}^{N} v_n(s)}. \tag{11}$$

Define $w(s) := [w_1(s), \cdots, w_N(s)]$ and let $b(s)$ be the influence rate; this is similar to the learning parameter $a(s)$ of the previous section and satisfies here, $b(s) \to 0$ as $s \to \infty$, $\sum_{s=1}^{\infty} b(s) = \infty$, and $\sum_{s=1}^{\infty} b^2(s) < \infty$. The influence of $\chi_n(s)$ on the user interest $\theta(s+1)$ is modeled as follows.

$$\theta(s+1) = (1 - b(s))\theta(s) + b(s)w(s). \tag{12}$$

Recall that the recommendation strategy $x(s)$ is also adapting using Eqn. (7), where $a(s)$ is adaptation parameter for RS. The evolution of $x(s)$ and $\theta(s)$ is of interest to us here and we would like to analyze the asymptotic behaviour of $x(s)$ and $\theta(s)$ as $s \to \infty$. The relation between the timescales at which $x(s)$ and $\theta(s)$ change determines the asymptotics and the relative timescales can be modeled by having $a(s)$ and $b(s)$ go to 0 at different rates. For example, if $b(s) \to 0$ much faster than $a(s)$ then this models $\theta(s)$ changing at a slower timescale than $x(s)$. The exact opposite of can also be modeled. Finally, we could also have both these change at the same rate and this is modeled with $a(s) = b(s)$.

The convergence analysis of $\theta(s)$ and $x(s)$ for different combinations of the relative timescales over which they change is given in Appendix G. The analysis is based on the two timescale stochastic approximation approach and the technique is detailed in [12, Chapter 6].

### A. Numerical Examples

We use the following parameter values. $r = [150, 125, 150, 400, 100]$, the initial user interest $\theta(1) = [0.25, 0.25, 0.25, 0.25]$, initial strategy $x(1) = [0.2, 0.2, 0.2, 0.2, 0.2]$, user characteristics for topics $\xi = [0.03, 0.05, 0.02, 0.9]$, and $\beta = [3, 3, 3, 3]$.

In Fig. IV-A, we show an example for the type-B influence model when $x(s)$ adapts slower than $\theta(s)$, i.e., $\frac{b(s)}{a(s)} \to \infty$. Specifically, $a(s) = \frac{0.1}{1+s}$ and $b(s) = \frac{0.5}{1+s^{0.6}}$. We observe that $x(s) \to \tilde{x}$, $\theta(s) \to \tilde{\theta}$ and $F(x(s), \theta(s)) \to F(\tilde{x}, \tilde{\theta})$ as $s$ increases. $\tilde{x} = [0.01, 0.01, 0.01, 0.96, 0.01]$,

$\tilde{\theta} = [0.002, 0.003, 0.005, 0.98]$ and $F(\tilde{x}, \tilde{\theta}) = 380$. The $\tilde{\theta}$ becomes more skewed towards topic 4 because (1) user intrinsic interest aligns towards topic 4, (2) the highest paying website 4 covers topic 4 with high frequency, and (3) the influence model is type-B, where user interest increases for repeatedly shown topics. This in turn leads to increase in the net revenue.

We next see the effect of relative timescales of influence on the type-B model in Fig. 3. Here, the dark line indicates the timescales $a(s) = \frac{0.1}{1+s}$, $b(s) = \frac{0.5}{1+s^{0.6}}$ and dotted line indicates the timescales $a(s) = \frac{0.1}{1+s^{0.6}}$, $b(s) = \frac{0.5}{1+s}$. In both these cases, the steady state value of $x(s) = [0.01, 0.01, 0.01, 0.96, 0.01]$. This is due to website 4 offer the highest payoff and type-B influence model. It is interesting to observe the effect of timescales on $\theta(s)$ and $F(x(s), \theta(s))$. When $\frac{b(s)}{a(s)} \to \infty$ as $s \to \infty$, the user interest $\theta(s)$ skews further towards topic 4 (exceeding 0.9 in 50 sessions). When $\frac{b(s)}{a(s)} \to 0$, as $s \to \infty$, the user interest $\theta(s)$ is skewed towards topic 4 and is less 0.9. Thus when user interest adapts faster than recommendation startegy, i.e., $\frac{b(s)}{a(s)} \to \infty$, it has a positive impact and leads to increase in net revenue compared to when $\frac{b(s)}{a(s)} \to 0$.

The preceding are just samples of how, the influence model, user intrinsic interest, influence factor, the publishing matrix, the relative timescales between recommendation learning rate $a(s)$, and influence rate $b(s)$ affect the determination of steady state values of $x(s)$, $\theta(s)$ and the net revenue. Additional numerical results are available in Appendix.

## V. Summary and Discussion

In this paper, we used a learning algorithm developed for a class of recommendation systems motivated by the `Stumbleupon` website and adapted it to the case when the RS receives a website-dependent reward for accepted recommendations. A key novelty of our work is that we also consider the case where the user interests are not 'rigid', but are themselves influenced by the recommendations made by the system.

Clearly, this is just a scratching of the surface. Extensive numerical investigation will help explain the phenomena better. It would also be good to develop such influence models for other online learning algorithms. While we would like some empirical validation of the models, we believe that would require design of suitable experiments which would be a separate line of inquiry.

## Appendix

### A. Derivation of Eqn. 3

$$C(x) = \sum_{n=1}^{N} \theta_n \sum_{l=1}^{\infty} c_l \rho_n (1-\rho_n)^{(l-1)}.$$

We can write $c_l = \sum_{k=1}^{l} \Delta c_l = (c_l - c_{l-1}) + \cdots + (c_1 - c_0)$, and $c_0 = 0$. Hence

$$C(x) = \sum_{n=1}^{N} \theta_n \sum_{l=1}^{\infty} \left( \sum_{k=1}^{l} \Delta c_k \right) \rho_n (1-\rho_n)^{(l-1)}.$$

We interchange summations and after some simplification we obtain

$$C(x) = \sum_{n=1}^{N} \theta_n \sum_{k=1}^{\infty} \Delta c_k \sum_{l=k}^{\infty} \rho_n (1-\rho_n)^{(l-1)}$$

After separating summations, we have

$$\begin{aligned} C(x) &= \sum_{n=1}^{N} \theta_n \left[ \sum_{k=1}^{\infty} \Delta c_k (1-\rho_n)^{(k-1)} \sum_{l=0}^{\infty} \rho_n (1-\rho_n)^{l} \right] \\ &= \sum_{n=1}^{N} \theta_n \left[ \sum_{k=1}^{\infty} \Delta c_k (1-\rho_n)^{(k-1)} \right] \end{aligned}$$

□

### B. Proof of Lemma 1

Using the definition of $\Delta c_l$ from (4), we can rewrite first term of $C(x)$ as follows.

$$C(x) = \sum_{n=1}^{N} \theta_n \left( \sum_{l=1}^{\infty} \Delta c_l \, (1-\rho_n)^{l-1} \right). \quad (13)$$

From our assumption that $c_l$ is convex increasing, $\Delta c_l \geq 0$ and is non decreasing in $l$ for all $l \geq 1$. Further, for a fixed $l > 1$, $(1-\rho_n)^{l-1}$ is a convex function of $\rho_n$ for $\rho_n \in [0,1]$ and for $l = 1$ it is a constant. Thus $\sum_{l=1}^{\infty} \Delta c_l \, (1-\rho_n)^{l-1}$ is also convex function of $\rho_n$ for $\rho_n \in (0,1]$. Note that $\rho_n > 0$. Since the RHS of Eqn. (13) is convex combination of convex functions of $\rho_n$, it is also a convex in $\rho_n$. Finally, from the definition of $\rho_n$ in (1), we see that $\rho_n$ is linear transformation of $x$ which in turn means that first term of $C(x)$, being a composition of convex function with a linear function transformation is also convex [?, Prop. 1.14]. This completes the proof. □

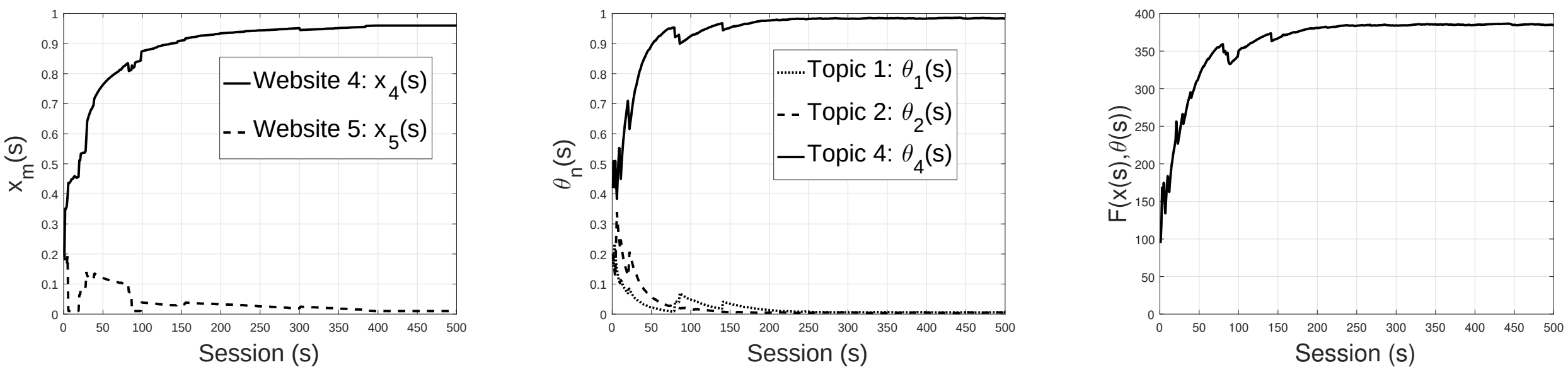

a) b) c)

Fig. 2. In plot (a), we plot the $x_m(s)$ as a function of $s$ for $m = 4, 5$. $\theta_n(s)$ for $n = 1, 2, 4$ is shown in plot (b). In plot (c), we illustrates the objective function $F$ as function of $s$. Here, we use influence type-B model.

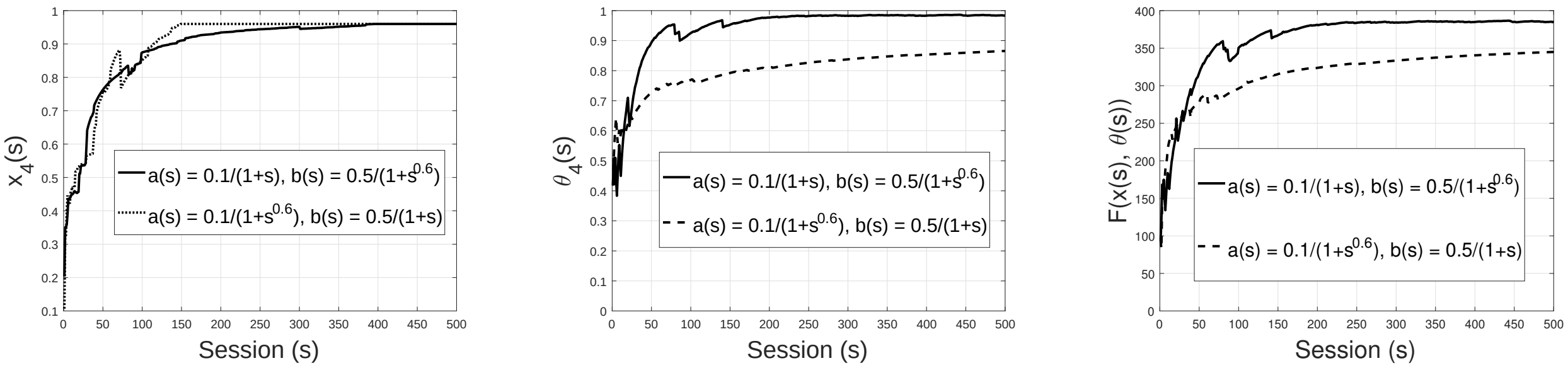

a) b) c)

Fig. 3. In plot (a), we describe $x_4(s)$ as function of session $s$. We illustrate $\theta_4(s)$ verses session $s$ in plot (b). Plot (c) illustrates the objective function $F$ as function of $s$. We use influence type-B model. Here, we see the effect of relative timescales between $x(s)$ and $\theta(s)$.

### C. Derivation of Eqn. (5)

We first derive $\frac{\partial R(x)}{\partial x_m}$. We rearrange $R(x)$ in the following way.

$$\begin{aligned} R(x) &= \sum_{n=1}^{N} \theta_n \frac{h_n(x)}{\rho_n(x)} \\ h_n(x) &= \sum_{m=1}^{M} r_m x_m p_{m,n} \\ \rho_n(x) &= \sum_{k=1}^{M} x_m p_{m,n}. \end{aligned}$$

We now take partial derivative of $R(x)$ w.r.t. $x_m$ and obtain

$$\begin{aligned} \frac{\partial R(x)}{\partial x_m} &= \sum_{n=1}^{N} \theta_n \left[ \frac{1}{\rho_n(x)} \frac{\partial h_n(x)}{\partial x_m} - \frac{h_n(x)}{(\rho_n(x))^2} \frac{\partial \rho_n(x)}{\partial x_m} \right] \\ &= \sum_{n=1}^{N} \theta_n \left[ \frac{r_m p_{m,n}}{\rho_n(x)} - p_{m,n} \frac{\sum_{k=1}^{M} x_k r_k p_{k,n}}{(\rho_n(x))^2} \right] \\ &= \sum_{n=1}^{N} \frac{\theta_n p_{m,n}}{\rho_n(x)} \left[ r_m - \frac{1}{\rho_n(x)} \sum_{k=1}^{M} x_k r_k p_{k,m} \right]. \end{aligned}$$

We next derive $\frac{\partial C(x)}{\partial x_m}$. Recall that $C(x)$ can be written as follows.

$$C(x) = \sum_{n=1}^{N} \theta_n \left[ \sum_{k=1}^{\infty} \Delta c_k (1 - \rho_n)^{(k-1)} \right].$$

Taking partial derivative of $C(x)$ w.r.t. $x_m$, we have

$$\begin{aligned} \frac{\partial C(x)}{\partial x_m} &= \sum_{n=1}^{N} \theta_n \left[ -\sum_{k=1}^{\infty} (k-1)(1-\rho_n)^{(k-2)} p_{m,n} \Delta c_k \right] \\ &= -\sum_{n=1}^{N} \theta_n p_{m,n} \left[ \sum_{l=0}^{\infty} l (1-\rho_n)^{(l-1)} \Delta c_{l+1} \right]. \end{aligned}$$

Second equiality because we substitute $k - 1 = l$. This gives us required result. □

### D. Proof of Lemma 2

Clearly, $R(x)$ abd $C(x)$ are differentiable function of $x$. Hence $\mathrm{F}(x)$ is differentiable function of $x$. To prove Lipschitz property, we use the mean value theorem.

1) From the mean value theorem, we can have

$$\mathrm{F}(\tilde{x}) - \mathrm{F}(\hat{x}) = \nabla \mathrm{F}(x)^T (\tilde{x} - \hat{x}),$$

where, $x = \zeta \tilde{x} + (1 - \zeta)\hat{x}$, $\zeta \in [0, 1]$ for $\tilde{x}, \hat{x} \in \mathcal{D}_\epsilon$. From Holder's inequality, we obtain

$$||\mathrm{F}(\tilde{x}) - \mathrm{F}(\hat{x})||_1 \le ||\nabla \mathrm{F}(x)^T||_\infty ||\tilde{x} - \hat{x}||_1.$$

By infinity norm property, we have

$$||\nabla \mathrm{F}(x)||_\infty \le ||\nabla R(x)||_\infty + ||\nabla C(x)||_\infty.$$

2) We first provide bound on $||\nabla R(x)||_\infty$. To obtain this upper bound, we require to provide bound on $\frac{\partial R(x)}{\partial x_m}$. We rewrite following from Eqn. 5.

$$\frac{\partial R(x)}{\partial x_m} = \sum_{n=1}^{N} \frac{\theta_n p_{m,n}}{\rho_n}\left[r_m - \sum_{k=1}^{M}\frac{x_k p_{k,n}}{\rho_n} r_k\right].$$

Notice that $r, x, \theta$ and $P$ are finite. Since $\rho_n \ge \delta > 0$, $\frac{1}{\rho_n}$ is finite for all $n$. Then we can have $|\frac{\partial R(x)}{\partial x_m}| < \infty$. Let $|\frac{\partial R(x)}{\partial x_m}| = L_{1,m} < \infty$. Thus $||\nabla R(x)||_\infty = \max_{1\le m\le M} L_{1,m} = L_1$ This holds for all $x \in \mathcal{D}_\epsilon$.

3) We now give bound on $||\nabla C(x)||_\infty$. Consider

$$\frac{\partial C(x)}{\partial x_m} = -\sum_{n=1}^{N}\theta_n p_{m,n}\left[\sum_{l=1}^{\infty} l\Delta c_{l+1}(1-\rho_n)^{l-1}\right].$$

Observe that $l\Delta c_{l+1}$ is a polynomial function because $c_l$ is a polynomial function. Further, from Eqn. 4, we can obtain $l\Delta c_{l+1} \le l^{(\kappa+1)}$, where $\kappa < \infty$. Also note that $(1-\rho_n)^{l-1}$ is exponetially decaying in $l$. Hence $\sum_{l=1}^{\infty} l\Delta c_{l+1}(1-\rho_n)^{l-1} < \infty$. Then we can have $|\frac{\partial C(x)}{\partial x_m}| < \infty$. Let $|\frac{\partial C(x)}{\partial x_m}| = L_{2,m}$. Thus

$$\begin{aligned} ||\nabla C(x)||_\infty &= \max_{1\le m\le M}\left|\frac{\partial C(x)}{\partial x_m}\right| \\ &\le \max_{1\le m\le M} L_{2,m} = L_2. \end{aligned}$$

This leads to bound on $||\nabla F(x)||_\infty$. Thus $||\nabla F(x)||_\infty \le L_1 + L_2 = \tilde{L}$. □

### *E. Preliminary Results Towards Proving Lemma 3*

We here show that $E[g_m(s)] = \frac{\partial F(x(s))}{\partial x_m(s)}$. This result will be used for showing that $g_m(s)$ is a noisy version of gradient of $F(x(s))$.

*Lemma 4:* Let $g_m(s)$ be as defined in Eqn. (6). Then we can obtain the following.

$$E[g_m(s)] = \frac{\partial F(x(s))}{\partial x_m(s)},$$

where

$$\frac{\partial F(x(s))}{\partial x_m(s)} = \frac{\partial R(x(s))}{\partial x_m(s)} - \frac{\partial C(x(s))}{\partial x_m(s)}.$$

*Proof:* Let $I$ be the event associated with the user interest in topic in a session. Let $J$ be the event that the site is click through in a session. We first obtain conditional probability of the site $m$ click through given that the user interest in topic $n$, it is given by

$$\Pr(J = m \mid I = n) = \frac{x_m p_{m,n}}{\sum_{m=1}^{M} x_m p_{m,n}}.$$

Recall that $T$ denoted number of trials that required for click through happens. Given that the user interest in topic $n$, the conditional probability that number of trials are $l$ is as follows.

$$\Pr(T = l \mid I = n) = \rho_n(1-\rho_n)^{(l-1)}.$$

We compute the joint conditional probability of site $m$ is clicked through in $l$th trial in a session given that the user interest is in topic $n$.

$$\begin{aligned} \Pr(J = m \text{ and } T = l \mid I = n) &= (1-\rho_n)^{(l-1)} x_m p_{m,n} \\ &= \rho_n(1-\rho_n)^{(l-1)}\frac{x_m p_{m,n}}{\rho_n}. \end{aligned}$$

Hence

$$\Pr(J = m \text{ and } T = l \mid I = n) = \Pr(J = m \mid I = n) \times \Pr(T = l \mid I = n).$$

We now ready to compute the expectation of $g_m(s)$. This is given as follows.

$$E[g_m(s)] = \sum_{n=1}^{N}\theta_n \sum_{l=1}^{L}\Pr(J = m \text{ and } T = l \mid I = n)\, g_m(s). \tag{14}$$

After substituting $g_m(s)$ from Eqn. (6) in previous Eqn. (14), and simplifying expressions, we obtain the desired result. ■

### *F. Proof of Lemma 3*

1) We here show that $E[V(x(s))] = 0$. From Eqn. (5), we can rewrite

$$\frac{\partial R(x(s))}{\partial x_m(s)} = \frac{1}{x_m(s)}\sum_{n=1}^{N}\theta_n \frac{x_m(s) p_{m,n}}{\rho_n} \times \left[r_m - \sum_{k=1}^{M}\frac{x_k(s) p_{k,n}}{\rho_n(s)} r_k\right],$$

$$\frac{\partial C(x(s))}{\partial x_m(s)} = -\frac{1}{x_m(s)}\sum_{n=1}^{N}\theta_n \frac{x_m(s) p_{m,n}}{\rho_n} \times \left(\sum_{l=1}^{\infty} l\Delta c_{l+1}\rho_n(1-\rho_n(s))^{l-1}\right).$$

From Lemma 4, we have

$$\begin{aligned} E[g_m(s)] &= \frac{\partial F(x(s))}{\partial x_m(s)} \\ &= \frac{\partial R(x(s))}{\partial x_m(s)} - \frac{\partial C(x(s))}{\partial x_m(s)}, \end{aligned}$$

and we obtain

$$E[V(x(s))] = \nabla F(x(s)) - E[G(s)] = 0.$$

Here, $E[G(s)] = [E[g_1(s)], \cdots, E[g_M(s)]]$.

2) We now want to show that $E[||V(x(s))||^2] < \infty$. We can write $E[||V(x(s))||^2]$ as follows.

$$\begin{aligned} E[||V(x(s))||^2] &= \sum_{m=1}^{M}\left[E[|g_m(s)|^2] - (\mathbb{E}g_m(s)])^2\right] \\ &\le \sum_{m=1}^{M} E[|g_m(s)|^2]. \end{aligned}$$

This suggest that we have to prove $E[|g_m(s)|^2] < \infty$. From Eqn. (6), we notice that $g_m(s)$ is addition of two term, and we use inequality $(A+B)^2 \le 2(A^2+B^2)$. Let

$$A := \frac{1}{x_m(s)}\left[r_m - \frac{1}{\rho_n(s)}\sum_{k=1}^{M} r_k x_k(s) p_{k,n}\right]$$
$$B := \frac{T(s)\Delta c_{T(s)+1}}{x_m(s)}.$$

We now write $E[A^2]$ as follows.

$$E[A^2] = \sum_{n=1}^{N}\theta_n \sum_{l=1}^{\infty}\rho_n(s)(1-\rho_n(s))^{(l-1)}\frac{x_m(s)p_{m,n}}{\rho_n(s)} \times \left(\frac{1}{x_m(s)}\left[r_m - \frac{1}{\rho_n(s)}\sum_{k=1}^{M} r_k x_k(s)p_{k,n}\right]\right)^2.$$

Rearranging terms, we get

$$E[A^2] = \sum_{n=1}^{N}\theta_n \frac{x_m(s)p_{m,n}}{\rho_n(s)} \times \left(\frac{1}{x_m(s)}\left[r_m - \frac{1}{\rho_n(s)}\sum_{k=1}^{M} r_k x_k(s)p_{k,n}\right]\right)^2.$$

Observe that all terms in the previous expression are finite, because $\rho_n(s) \ge \delta > 0$, $x_m(s) \ge \epsilon$, $r, P$, and $x(s)$ are finite. Further it is sum of finite terms. Hence $E[A^2] < \infty$.
We now want to bound $E[B^2]$. We can write

$$E[B^2] = \sum_{n=1}^{N}\theta_n \sum_{l=1}^{\infty}\rho_n(s)(1-\rho_n(s))^{(l-1)} \times \frac{x_m(s)p_{m,n}}{\rho_n}\left(\frac{l\Delta c_{l+1}}{x_m(s)}\right)^2.$$

After simplifications, we have

$$E[B^2] = \sum_{n=1}^{N}\theta_n \frac{p_{m,n}}{\rho_n(s)}\frac{1}{x_m(s)}\sum_{l=1}^{\infty}(l\Delta c_{l+1})^2 \times \rho_n(s)(1-\rho_n(s))^{(l-1)}.$$

Note that we considered $c_l = l^{\kappa}$, this implies that $l\Delta c_{l+1} \le l^{(\kappa+1)}$ and $(l\Delta c_{l+1})^2 \le l^{2(\kappa+1)}$. It is polynomial in $\kappa$. $(1-\rho_n(s))^2$ decays exponentially in $l$. Hence

$$\sum_{l=1}^{\infty}(l\Delta c_{l+1})^2 \rho_n(s)(1-\rho_n(s))^{(l-1)} < \infty, \quad (15)$$

and $E[B^2] < \infty$.

3) Thus from the preceding results, we obtain $E[||V(x(s))||^2] < \infty$. This completes the proof. □

### G. Analysis for convergence of $x(s)$ and $\theta(s)$

In the following we sketch the key steps involved in the analysis. We first analyze scenario when a user adapts on much slower timescales than the recommendation strategy of RS, $\theta(s)$ changes on a slower timescale than $x(s)$. The key steps for analysis are the following.

1) We have already assumed that $a(s)$ and $b(s)$ satisfy
$$\sum_{s=1}^{\infty} a(s) = \sum_{s=1}^{\infty} b(s) = \infty.$$
Further, to model $\theta(s)$ changing on a slower timescale than $x(s)$, we also assume $\frac{b(s)}{a(s)} \to 0$, $s \to \infty$. In addition we will also make the following technical assumption.
$$\sum_{s=1}^{\infty} a^2(s) + b^2(s) < \infty.$$
2) Like in the analysis of the online learning algorithm in Section III, we have to obtain the suitable limiting o.d.e. for the iterations of (7) and (12). Since these iterations are running at different timescales, we compare them to the solution of the singularly perturbed o.d.e.
$$\dot{x}(s) = \frac{1}{\gamma}\left(\nabla F(x(s),\theta(s)) + z\right) \quad (16)$$
$$\dot{\theta}(s) = h(x(s),\theta(s)). \quad (17)$$
as $\gamma \to 0$. $h$ is the expression of limiting o.d.e. for $\theta$.
3) This interacting o.d.e. is analyzed by assuming the slow component, $\theta(s)$, to be quasi-static, i.e. let $\theta(s) = \theta$. This reduces the o.d.e. of the fast component, $x(s)$, to have $\theta(s) = \theta$ as a parameter. Then we can analyze the solution of the latter o.d.e. using the standard approach. Fixing this solution of $x(s)$, we can then analyze the o.d.e. for $\theta(s)$. We thus have
$$\dot{x}(s) = \nabla F(x(s),\theta(s)) + z. \quad (18)$$
The solution of this o.d.e. is function $\theta$, say, $\lambda(\theta)$. For small $\gamma$, the iterate $x(s)$ closely tracks the solution of o.d.e. $\dot{x}(s)$, i.e., $\lambda(\theta)$. Hence we can have
$$\dot{\theta}(s) = h(\lambda(\theta(s)),\theta(s)). \quad (19)$$
Then we can show that $\theta(s)$ tracks the solution of o.d.e (19).
The procedure for analysis of the convergence of iterates (7) and (12) is to construct the piecewise linear and continuous interpolated processes. By assuming $\theta(s)$ to be quasi-static, we can show that this continuous interpolation of (7) converges to the solution of the o.d.e (18), i.e. to $\lambda(\theta)$. Fixing $\lambda(\theta)$, we can then show that the continuous interpolated process of iteration (12) tracks the solution of o.d.e. (19) and converges to $\theta^*$, the solution.
It can be shown that $(x(s),\theta(s)) \to (\lambda(\theta^*),\theta^*)$ almost surely. This result is shown in [12, Chapter 6, Theorem 2].

*Remark 2:* 1) We do not have close form expression for $h(x, \theta)$ but we believe that this can be obtained.

2) When $\theta(s)$ changes on a faster timescale than the recommendation strategy $x(s)$, then the analysis is identical to what we described above except that $\theta(s)$ and $x(s)$ get interchanged.

3) When the $\theta(s)$ and $x(s)$ change at the same rate, we assume $a(s) = b(s)$, and the convergence analysis of algorithm for $x(s)$ and $\theta(s)$ boils down to a single timescale stochastic approximation algorithm which is similar to the analysis of the previous section. This is summarised below. First observe that the iterations will become

$$\begin{bmatrix} x(s+1) \\ \theta(s+1) \end{bmatrix} = \Pi_{\mathcal{D}} \left[ \begin{bmatrix} x(s) \\ \theta(s) \end{bmatrix} + a(s) \begin{bmatrix} G(s) \\ w(s) - \theta(s) \end{bmatrix} \right]. \quad (20)$$

Here, $\Pi$ is Euclidean projection and $\mathcal{D}$ is suitably defined constraint set. Let

$$y(s+1) = \begin{bmatrix} x(s+1) \\ \theta(s+1). \end{bmatrix}$$

Then we can define constraint set as follows.

$$\mathcal{D} := \Bigg\{ y : y := [y_1, \cdots, y_{M+N}], \sum_{i=1}^{M} y_i = 1, \quad y_i \geq \epsilon > 0,$$

$$1 \leq i \leq M, \sum_{j=1}^{N} y_{M+j} = 1, y_j \geq 0, M+1 \leq j \leq M+N \Bigg\}.$$

Again using similar procedure in Section III, we can show that the iterations of (20) converge to the solution of the following limiting o.d.e.

$$\dot{y}(s) = \Phi(y(s)) + z,$$

where

$$\Phi(y(s)) = \begin{bmatrix} \nabla F(x(s), \theta(s)) \\ h(x(s)), \theta(s)) \end{bmatrix}.$$

### H. Static optimization: Additional numerical examples

*1) Example 3:* We consider the example with the highest paying website covers topics that match the user interest. We use $r = [150, 125, 150, 400, 100]$ and user interest $\theta = [0.1, 0.05, 0.25, 0.6]$. The optimal recommendation strategy is $x^* = [0.01, 0.01, 0.22, 0.66, 0.1]$ and corresponding net revenue is $F(x^*) = 253.2$. The $x^*$ recommends websites $\{3, 4, 5\}$ most frequently compare to other sites. The site 4 is the most frequently recommended among all sites because (1) it covers topics that match user interests, and (2) it has the highest payoff.

*2) Example 4:* We now consider the example, where the highest paying website cover topics that do not match the user interest. We use $r = [150, 125, 150, 400, 100]$ and user interest $\theta = [0.9, 0.05, 0.02, 0.03]$. The optimal recommendation strategy $x^* = [0.32, 0.01, 0.01, 0.29, 0.37]$ and corresponding net revenue $F(x^*) = 94$ is derived. The optimal net revenue $F(x^*)$ is less than preceding example. The $x^*$ recommends websites $\{1, 4, 5\}$ most frequently compare to other sites. Observe that the site 5 is the most frequently recommended site among all site because it covers topics that match user interests, even though there is the least payoff from website 5.

### I. Online learning: Additional numerical examples

In Fig. 4, we consider the example with user having different interest for different topics and different payoff from websites. The user interest is skewed towards topic 4 and reward structure is also skewed for website 4. We use $r = [150, 125, 150, 400, 100]$ and $\theta = [0.03, 0.05, 0.02, 0.9]$. We now observe from Fig. 4 that the recommendation strategy $x(s)$ converges to $x^*$, i.e., $x_3^* \to 0.25$, $x_4(s) \to 0.72$ and $x_5(s) \to 0.01$, and other $x_m(s) \to x_m^*$. See Table I, in that 3rd row is optimal $x^*$ and $x(s) \to x^*$ as $s \to \infty$. The convergence of $x(s)$ to one of the optimal $x^*$ is determined by the stepsizes $a(s)$. The optimal net revenue $F(x(s)) \to F(x^*)$. We also see that the error function is decreasing.

In Fig. 5, we describe the convergence of learning algorithm $x(s)$ for example-3 in Section H, where user interest is skewed towards topic 4 than other topics. The highest paying website 4 cover the topic that match the user interests. Once again we observe that $x_3(s) \to 0.08$, $x_4(s) \to 0.65$, and $x_5(s) \to 0.25$. The error function $||x(s) - x^*||$ is decreasing with increase of $s$.

We now illustrate the convergence of learning algorithm $x(s)$ for example-4 in Section H, where user interest is skewed towards topic 1 than other topics and the highest paying website 4 do not cover the topic that match the user interests. Once again we observe that $x_1(s) \to 0.32$, $x_4(s) \to 0.29$, and $x_5(s) \to 0.37$. The error function $||x(s) - x^*||$ is decreasing with increase of $s$. This demonstrate the convergence of our learning algorithm.

### J. Influencing user interests: Additional numerical examples

In Fig. 7 and 8, we consider that the $x(s)$ is updated at fast timescales and $\theta(s)$ is updated at slow timescales. We use $a(s) = \frac{0.001}{(1+s)}$, and $b(s) = \frac{0.9}{(1+s^{1.4})}$. Other parameters are as follows. The reward from websites $r = [150, 125, 150, 400, 100]$, the initial user interests $\theta(1) = [0.25, 0.25, 0.25, 0.25]$, initial recommendation strategy $x(1) = [0.2, 0.2, 0.2, 0.2, 0.2]$, a user characteristics for topics $\xi = [0.1, 0.05, 0.25, 0.6]$, and $\beta = [3, 3, 3, 3]$.

We illustrate a numerical example for the type-A influence model in Fig. 7,and we obtain $x(s) \to \hat{x}$, $\theta(s) \to \hat{\theta}$, and $F(x(s), \theta(s)) \to F(\hat{x}, \hat{\theta})$ as $s$ increases. Here, $\hat{x} = [0.03, 0.01, 0.31, 0.48, 0.17]$, $\hat{\theta} = [0.14, 0.09, 0.3, 0.47]$, and $F(\hat{x}, \hat{\theta}) = 200$. Since $\theta(s)$ has very slow learning rate, it quickly converges to $\hat{\theta}$.

In Fig. 8, we have numerical example for the type-B influence model and we obtain $x(s) \to \tilde{x}$, $\theta(s) \to \tilde{\theta}$, and $F(x(s), \theta(s)) \to F(\tilde{x}, \tilde{\theta})$ as $s$ increases. Here, $\tilde{x} = [0.03, 0.01, 0.3, 0.45, 0.2]$, $\tilde{\theta} = [0.1, 0.07, 0.21, 0.62]$, and $F(\tilde{x}, \tilde{\theta}) = 247$. Here also, $\theta(s)$ converges quickly to $\tilde{\theta}$.

From Fig. 7 and 8, we observe that user interests in steady state is different for both type-A and type-B model and it is more skewed for the type-B model. Hence the type-B influence model has higher net revenue in steady state compare to that of the type-A model.

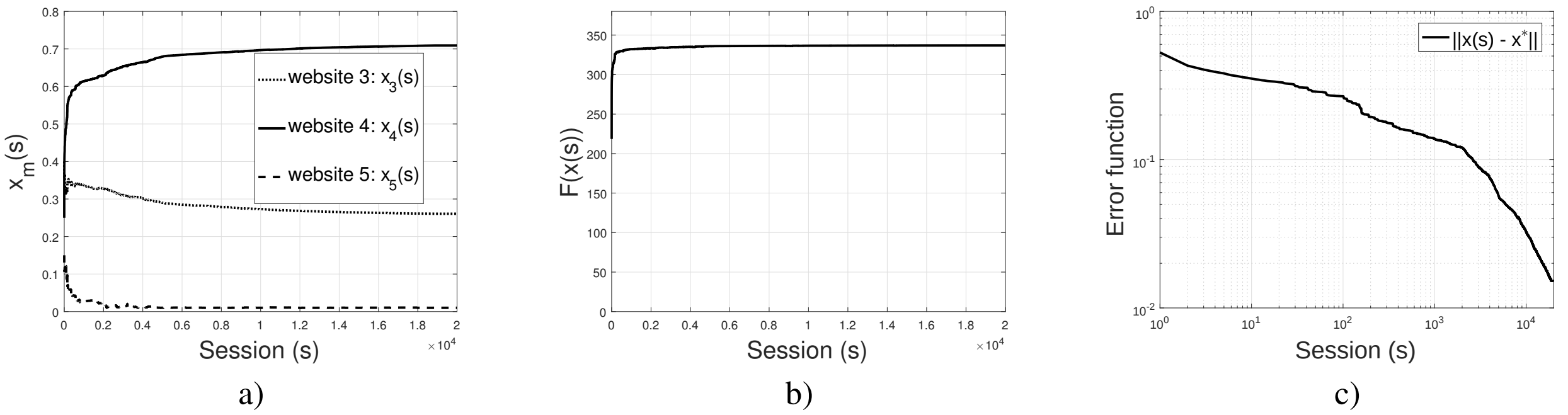

Fig. 4. Plot (a) illustrates the recommendation strategy $x_m(s)$ for website $m$ verses session $s$. Plot (b) describes the objective function $F$ verses session $s$. In plot (c), we describe the error function $||x(s) - x^*||$ as function of $s$. It is illustrated for a user that has different interest for different topics, has highest interests in topic 4 and reward structure from website is also skewed towards website 4. Here, we used $r = [150, 125, 150, 400, 100]$ and $\theta = [0.03, 0.05, 0.02, 0.9]$.

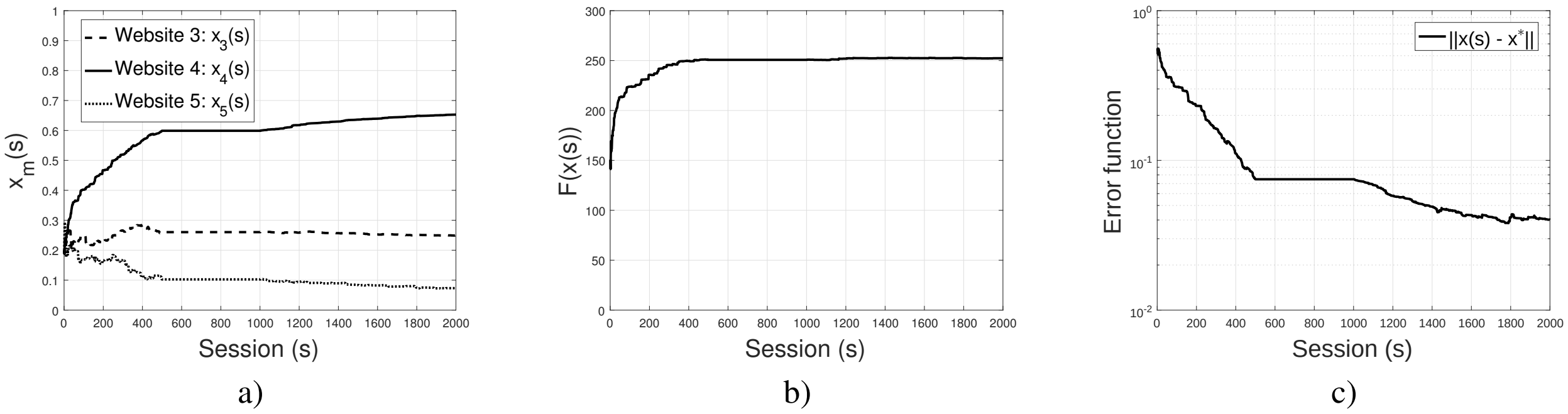

Fig. 5. Plot (a) illustrates the recommendation strategy $x_m(s)$ for website $m$ verses session $s$. Plot (b) describes the objective function $F$ verses session $s$. In plot (c), we describe the error function $||x(s) - x^*||$ as function of $s$. It is illustrated for a user that has different interest for different topics, has highest intersts in topic 4 and reward structure from website is also skewed towards website 4. Here, we used $r = [150, 125, 150, 400, 100]$ and $\theta = [0.1, 0.05, 0.25, 0.6]$.

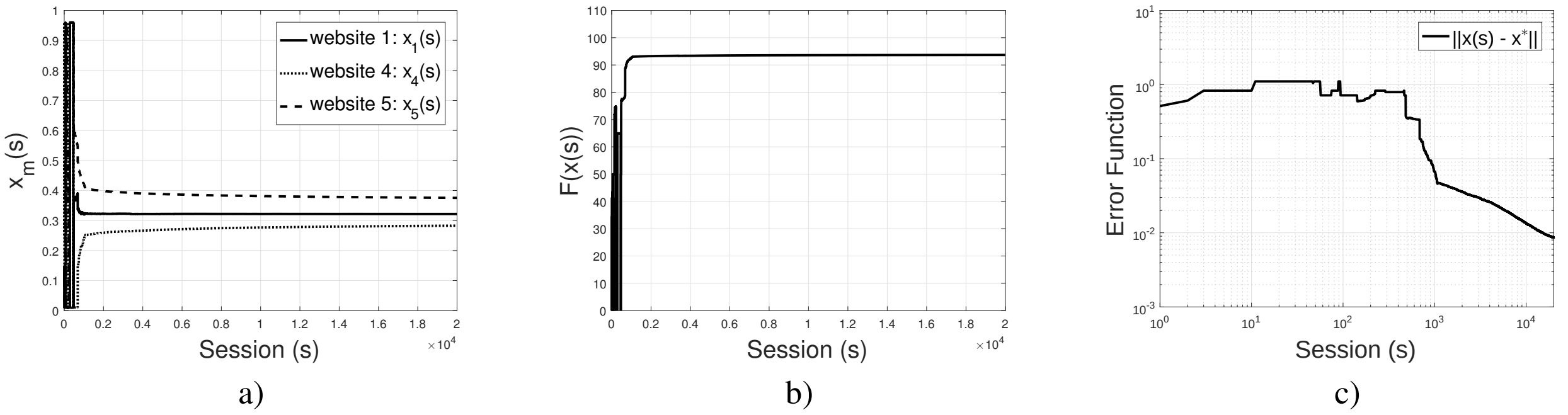

Fig. 6. Plot (a) illustrates the recommendation strategy $x_m(s)$ for website $m$ verses session $s$. Plot (b) describes the objective function $F$ verses session $s$. In plot (c), we describe the error function $||x(s) - x^*||$ as function of $s$. It is illustrated for a user that has skewed interest for topics and reward structure from website is also skewed. But high payoff website do not cover topics that matches the user interest. We used $r = [150, 125, 150, 400, 100]$ and $\theta = [0.9, 0.05, 0.02, 0.03]$.

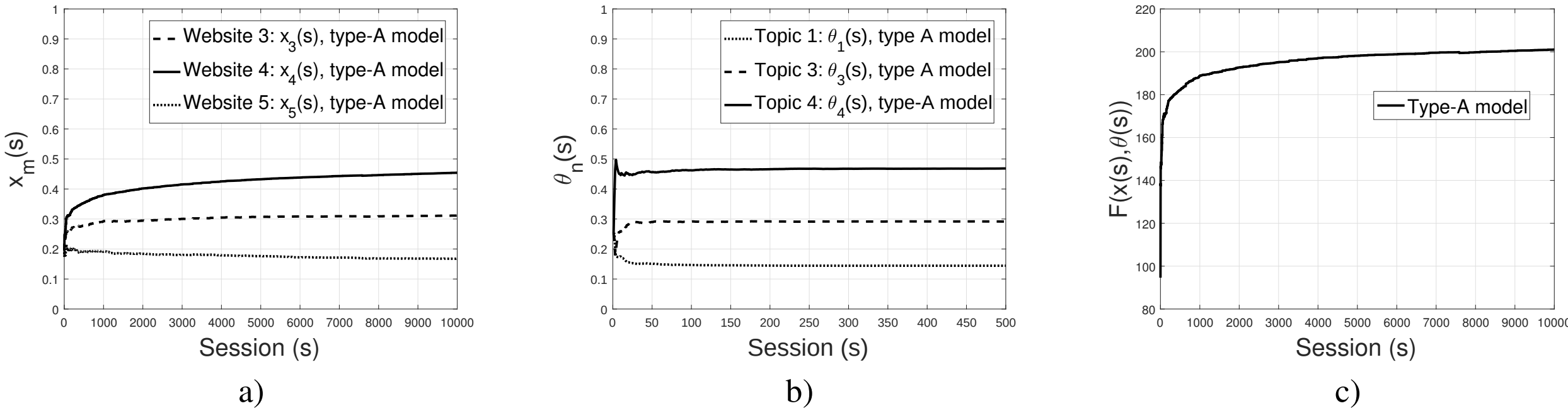

Fig. 7. In plot (a), we describe $x_m(s)$ as function of session $s$. We illustrate $\theta_n(s)$ verses session $s$ in plot (b). Plot (c) illustrates the objective function $F$ as function of $s$. We use type-A influence model. The $x(s)$ is updated with fast timescales and $\theta(s)$ is updated with slow timescales. Here, $r = [150, 125, 150, 400, 100]$ and $\xi = [0.1, 0.05, 0.25, 0.6]$.

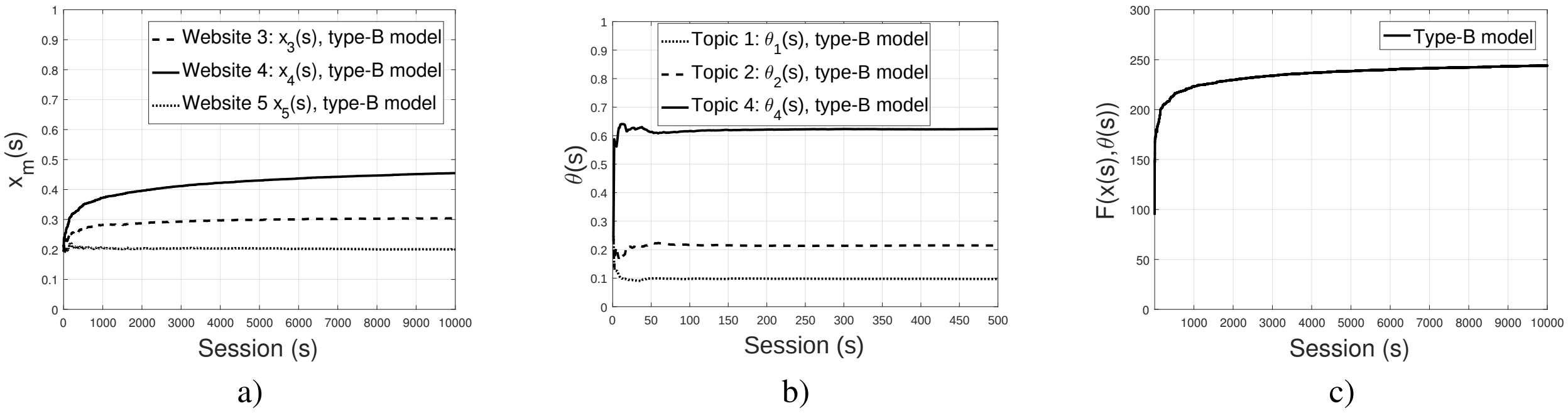

Fig. 8. In plot (a), we describe $x_m(s)$ as function of session $s$. We illustrate $\theta_n(s)$ verses session $s$ in plot (b). Plot (c) illustrates the objective function $F$ as function of $s$. We use type-B influence model. The $x(s)$ is with fast timescales and $\theta(s)$ is with slow time scales. Here, $r = [150, 125, 150, 400, 100]$ and $\xi = [0.1, 0.05, 0.25, 0.6]$.

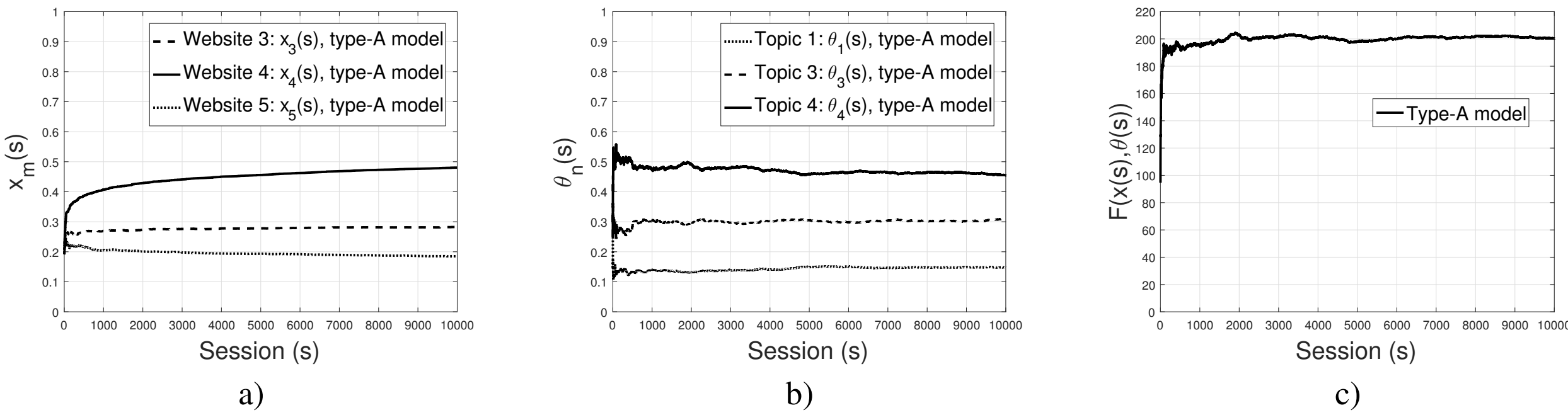

Fig. 9. In plot (a), we describe $x_m(s)$ as function of session $s$. We illustrate $\theta_n(s)$ verses session $s$ in plot (b). Plot (c) illustrates the objective function $F$ as function of $s$. We use type-A influence model. The $x(s)$ is updated with slow timescales and $\theta(s)$ is updated with fast timescales. Here, $r = [150, 125, 150, 400, 100]$ and $\xi = [0.1, 0.05, 0.25, 0.6]$.

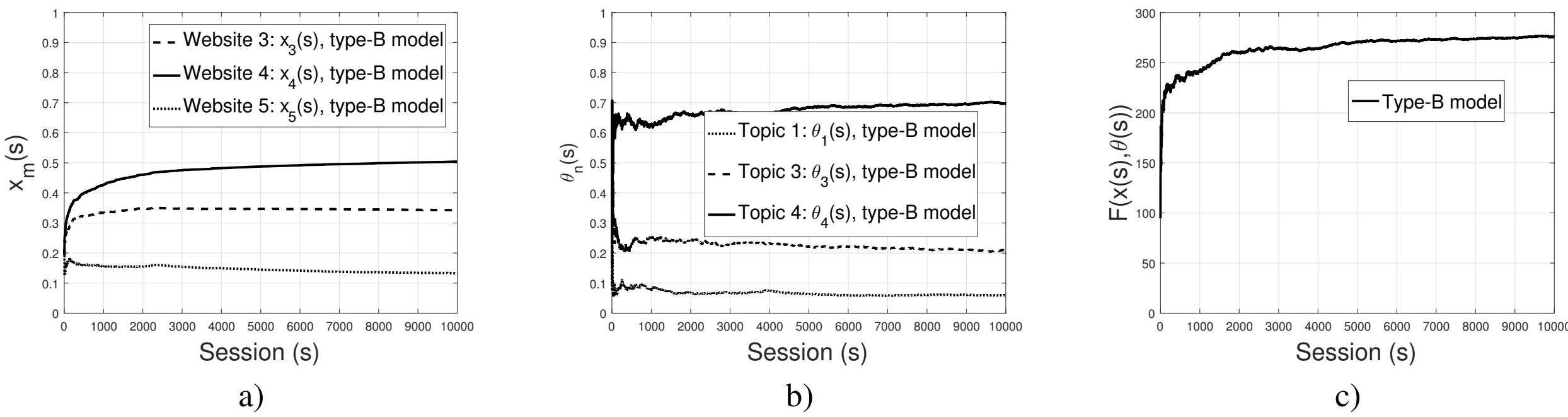

Fig. 10. In plot (a), we describe $x_m(s)$ as function of session $s$. We illustrate $\theta_n(s)$ verses session $s$ in plot (b). Plot (c) illustrates the objective function $F$ as function of $s$. We use type-B influence model. The $x(s)$ is with slow timescales and $\theta(s)$ is with fast timescales. Here, $r = [150, 125, 150, 400, 100]$ and $\xi = [0.1, 0.05, 0.25, 0.6]$.

We next consider the examples where $x(s)$ is adapted at slower timescales and $\theta(s)$ is adapted at faster timescales for both type-A and type-B model. This is illustrated in Fig. 9 and 10. Here, $a(s) = \frac{0.001}{1+s}$, $b(s) = \frac{0.9}{1+s^{0.8}}$. $r = [150, 125, 150, 400, 100]$, and $\xi = [0.1, 0.05, 0.25, 0.6]$.

We also notice from Fig. 8 and 10 that for the type-B model, the fast timescale adaptation of $\theta(s)$ yields the higher net revenue compare to slow timescale adaptation of $\theta(s)$. This is because the steady state value of user interest is higher for topic $4$ in fast timescale model compared to slow timescale model. The topic $4$ is covered by website $4$.

But for the type-A model, the relative change of timescales in $\theta(s)$ and $x(s)$ do not provide any gain in steady state value of the net revenue. This can be observed from Fig. 7 and 9.